%% file: main.tex
\documentclass[10pt,conference]{IEEEtran}
\IEEEoverridecommandlockouts
\usepackage{cite}
\usepackage{amsmath,amssymb,amsfonts}
\usepackage{graphicx}
\usepackage{textcomp}
\usepackage{xcolor}
\usepackage{booktabs}   
\usepackage{multirow}   
\usepackage{tabularx}   
\usepackage{url}
\usepackage{subcaption}
\usepackage{algorithm}
\usepackage{algpseudocode}  
\usepackage{array} 

\usepackage{tcolorbox}

\usepackage[colorlinks,bookmarksopen,bookmarksnumbered,citecolor=red, urlcolor=black]{hyperref}

\def\BibTeX{{\rm B\kern-.05em{\sc i\kern-.025em b}\kern-.08em
    T\kern-.1667em\lower.7ex\hbox{E}\kern-.125emX}}

\input{setup}

\begin{document}

\title{Adaptive Request Scheduling for CodeLLM Serving with SLA Guarantees}


\author{
    \IEEEauthorblockN{Shi Chang\IEEEauthorrefmark{2}, Boyuan Chen\IEEEauthorrefmark{1}, Kishanthan Thangarajah\IEEEauthorrefmark{1}, Hanan Lutfiyya\IEEEauthorrefmark{2}, Ahmed E. Hassan\IEEEauthorrefmark{3}}
    \IEEEauthorblockA{\IEEEauthorrefmark{1}Centre for Software Excellence, Huawei Canada}    
    \IEEEauthorblockA{\IEEEauthorrefmark{2}Western University, Canada, \IEEEauthorrefmark{3}Queen's University, Canada} 
    \IEEEauthorblockA{\texttt{\{schan543, hlutfiyy\}@uwo.ca, ahmed@cs.queensu.ca, cse@huawei.com}}
}

\maketitle

\begin{abstract}

Code Large Language Models (CodeLLMs) are increasingly integrated into modern software development workflows, yet efficiently serving them in resource-constrained, self-hosted environments remains a significant challenge. Existing LLM serving systems employs Continuous Batching for throughput improvement. However, they rely on static batch size configurations that cannot adapt to fluctuating request rates or heterogeneous workloads, leading to frequent SLA (Service Level Agreement) violations and unstable performance. In this study, We propose~\textsc{SABER}, a dynamic batching strategy that predicts per-request SLA feasibility and adjusts decisions in real time. \textsc{SABER} improves goodput by up to 26\% over the best static configurations and reduces latency variability by up to 45\%, all without manual tuning or service restarts. Our results demonstrate that SLA-aware, adaptive scheduling is key to robust, high-performance CodeLLM serving.

\end{abstract}


\section{Introduction}
\label{sec:intro}
\input{sections/1.introduction}

\input{sections/2.background}


\input{sections/3.motivation}

\input{sections/4.system}

\input{sections/5.RQ1}

\input{sections/6.RQ2}

\input{sections/7.discussion}

\input{sections/8.theats}

\input{sections/9.conclusion}

\input{sections/10.disclaimer}

\clearpage
\bibliographystyle{IEEEtran}
\bibliography{main}

\end{document}

%% file: setup.tex
\usepackage{xspace}
\usepackage{soul}
\setuldepth{Berlin}
\usepackage{hyperref}
\hypersetup{colorlinks=true,breaklinks=true}
\usepackage{multirow}
\usepackage{makecell}
\usepackage{balance}
\usepackage{enumitem}
\usepackage{pifont}
\usepackage{float}
\usepackage{csquotes}

\newboolean{showcomments}
\setboolean{showcomments}{true}

\ifthenelse{\boolean{showcomments}}
{\newcommand{\nbc}[3]{
 {\colorbox{#3}{\bfseries\sffamily\scriptsize\textcolor{white}{#1}}}
 {\textcolor{#3}{\sf\small$\blacktriangleright$\textit{#2}$\blacktriangleleft$}}
 }
}
{\newcommand{\nbc}[3]{}
 }

%% file: sections/1.introduction.tex
In modern software development, Code Large Language Models (CodeLLMs)~\cite{codellama,deepseek-coder,hui2024qwen2, starcoder},have rapidly gained popularity for multiple coding tasks~\cite{li2024fine,phan2024repohyper,jiang2024survey,izadi2024language}, particularly among indie developers and small development teams~\cite{murali2024aiassistedcodeauthoringscale}.
Due to privacy concerns, limited budgets, and customization needs, resource-constrained indie developers and small teams frequently choose to serve fine-tuned CodeLLMs locally or via rented acceleration units (e.g GPU, NPU etc) rather than relying on external model APIs~\cite{yao2024survey,das2025security,dhulshette2025hierarchical}.
However, self-hosting introduces significant operational challenges: developers must maintain high throughput and stable latency without dedicated infrastructure teams or specialized expertise.

Several serving frameworks attempt to simplify CodeLLM deployment: Ollama~\cite{ollama} makes model deployment as simple as installing an application; vLLM~\cite{vllm} significantly improves throughput by efficient memory management; SGLang~\cite{sglang} focuses on parallel execution of complex tasks. These frameworks employ continuous batching~\cite{daniel2023continuous, yu2022orca}, a technique that allows incoming requests to join executing batches immediately rather than waiting for the current batch to complete, thereby improving acceleration unit utilization and system throughput. However, enhanced throughput usually comes with the costs of request latency, in particular, strict latency Service-Level Agreements (SLAs) are crucial for coding tasks. Delays of response can rapidly degrade developer productivity and satisfaction.

Therefore, a batch size needs to be set to control the maximum number of requests can be processed in parallel. This configuration directly impacts system throughput and request latency~\cite{llumnix}. Yet, a static batch size configuration needs to be set by developers at system startup, (e.g \texttt{max\_seq} in vLLM~\cite{vllm}, \texttt{batch\_size} in Ollama~\cite{ollama}), which can be suboptimal because: a), workloads vary substantially throughout the day, shifting rapidly between shorter tasks like code summarization and lengthier tasks like full code generation or translation, leading to considerable variability in end-to-end request latencies. And b), to adapt batch size settings in response to workload changes, developers must frequently restart the serving engine, incurring downtime and unnecessary operational costs.

To address these limitations, we introduce~\textsc{SABER}, an automated, dynamic, and SLA-aware batching strategy designed for continuous batching model serving systems. \textsc{SABER} adaptively adjusts the batch size at runtime by predicting the likelihood of SLA compliance for incoming requests, proactively prioritizing viable requests, and deferring those unlikely to meet their deadlines. Importantly, this adjustment occurs on-the-fly without requiring developers to interrupt or restart the serving engine, significantly reducing operational overhead and manual intervention.

We empirically validate~\textsc{SABER} using realistic CodeLLM workloads under varying request loads and compositions, comparing its performance against the optimal static configuration for each scenario. Our results consistently demonstrate that~\textsc{SABER} achieves superior goodput and significantly more predictable latency, especially under conditions of high resource contention.

In summary, our contributions include: 
\begin{itemize}
    \item We identify and discuss the shortcomings of static continuous batching in realistic CodeLLM serving scenarios,
    \item We propose \textsc{SABER}, a novel batching approach designed specifically for CodeLLM workloads, and
    \item We provide the first empirical evaluation of continuous batching strategies from the perspective of latency SLAs and developer experience. 
\end{itemize}
To the best of our knowledge, this work represents the first systematic investigation into optimizing continuous batching configurations dynamically, directly improving both the performance and practical usability of self-hosted CodeLLM services.

The remainder of this paper is structured as follows: Section~\ref{sec:background} presents the background and related work. Section~\ref{sec:motivation} analyzes the fundamental limitations of static configurations through a motivational experiment. Section~\ref{sec:approach} describes the detailed design of~\textsc{SABER}. Sections~\ref{sec:rq1}  and Section~\ref{sec:rq2}  present experimental results demonstrating \textsc{SABER}'s effectiveness. Section~\ref{sec:discussion} discusses implications and limitations. Finally, Section \ref{sec:conclusion} concludes the paper.

%% file: sections/2.background.tex
\begin{figure*}[h]
    \centering
    \begin{subfigure}[b]{0.32\textwidth}
        \centering
        \includegraphics[width=\textwidth]{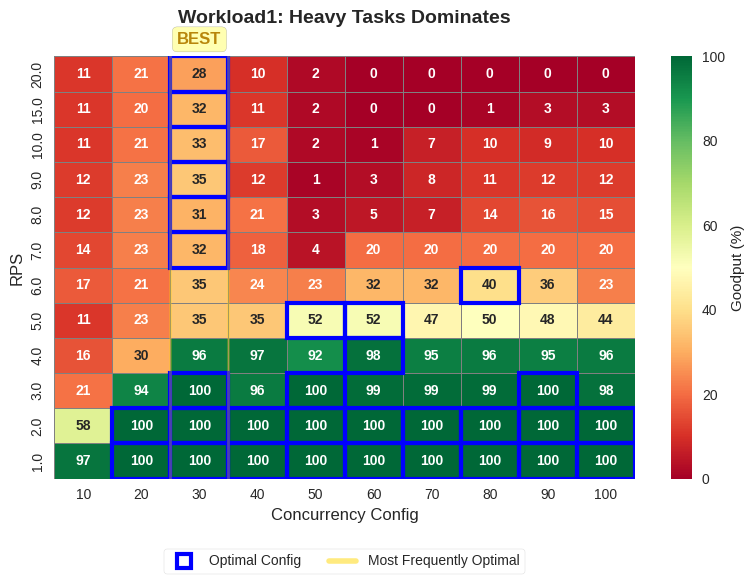}
        \caption{W1: Heavy Tasks}
        \label{fig:workload1_heatmap}
    \end{subfigure}
    \hfill
    \begin{subfigure}[b]{0.32\textwidth}
        \centering
        \includegraphics[width=\textwidth]{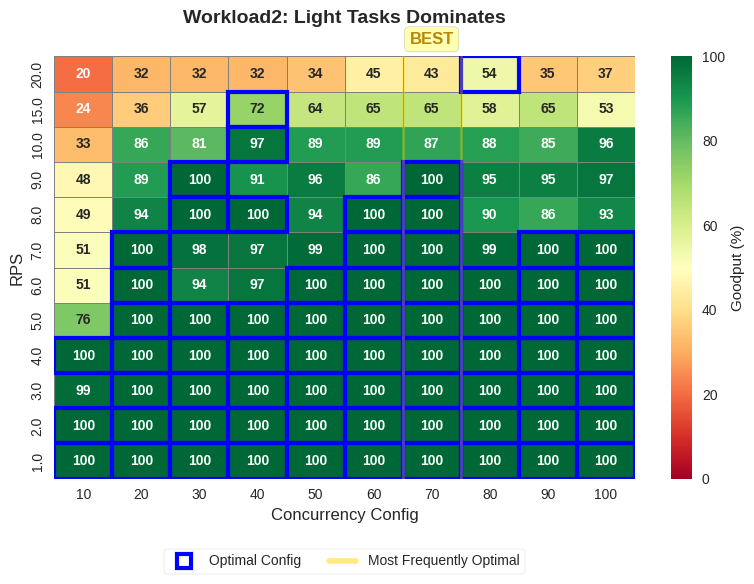}
        \caption{W2: Light Tasks}
        \label{fig:workload2_heatmap}
    \end{subfigure}
    \hfill
    \begin{subfigure}[b]{0.32\textwidth}
        \centering
        \includegraphics[width=\textwidth]{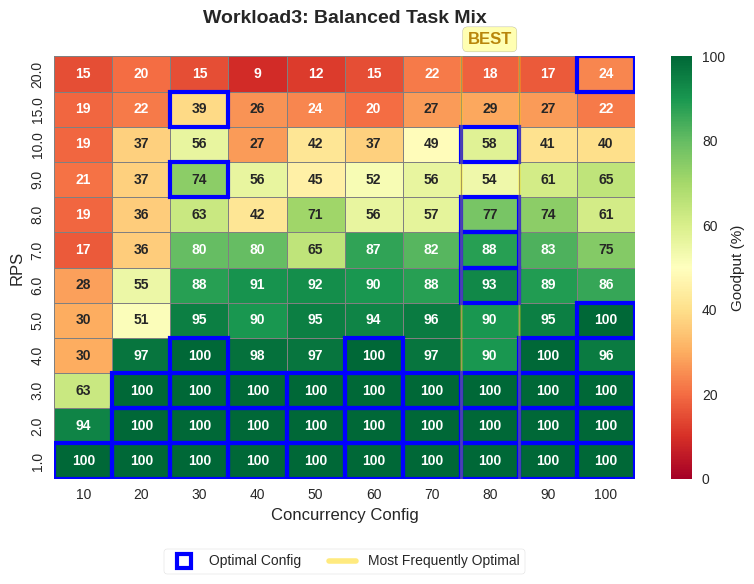}
        \caption{W3: Balanced Mix}
        \label{fig:workload3_heatmap}
    \end{subfigure}
    \caption{Goodput across different batch sizes and request rates.
    Each heatmap shows the percentage of requests meeting their SLA requirements (goodput). Green cells indicate high goodput (near 100\%), while red cells indicate low goodput.  
    Blue boxes mark the best configuration per RPS; the gold ``BEST" label marks the configuration that is best for the largest number of RPS values.}
    \label{fig:rq1_results}
\end{figure*}
\section{Background and Related Work}
\label{sec:background}
We first describe the background of employing request batching in Deep Neural Network (DNN) serving in Section~\ref{sec:background:batching} and then discuss the related work in Section~\ref{sec:background:related}.

\subsection{Static and Continuous Batching in DNN Model Serving}
\label{sec:background:batching}

Batching is a fundamental technique in deep neural network (DNN) serving~\cite{ali2020batch,cui2022dvabatch,yan2018efficient}, combining multiple inference requests into a single forward pass to maximize GPU utilization and amortize memory and compute overhead. For conventional architectures such as MLPs and CNNs, batching is straightforward: all requests in a batch are processed together with batched tensor operations. This works well because the computation is the same across requests: for instance, a single forward pass through a fixed-size image~\cite{terven2023comprehensive}.

However, large language models (LLMs), majority of which are decoder-only transformers, introduce new challenges for batching. In these models, inference is auto-regressive: the model generates one token at a time, and each request may require a different number of decoding steps depending on the prompt and stopping condition~\cite{he2025survey, cheng2024enabling}. This variability causes traditional static batching to perform poorly: short requests must wait for longer ones to finish, leading to idle compute and inflated tail latency.

To address this, modern serving engines adopt continuous batching~\cite{yu2022orca}, also known as iteration-level scheduling. In this approach, the batch is updated every decoding step: new requests can enter the batch mid-generation, and completed requests can exit immediately without waiting for others. This enables much higher throughput and GPU utilization by maintaining a saturated compute pipeline, even when requests have diverse lengths and arrival times.

Serving systems such as vLLM~\cite{vllm}, Ollama~\cite{ollama}, and SGLang~\cite{sglang} have widely adopted continuous batching as a default execution strategy. A key parameter in these systems is maximum batch concurrency, which limits the number of active decoding streams (i.e., concurrent requests) at each iteration. This setting is crucial for balancing throughput and latency, but it introduces new challenges: the level of interference a request experiences depends heavily on which other requests are co-scheduled in the same batch window. Developers have frequently reported performance unpredictability caused by suboptimal batching configurations~\cite{vllm10709,vllm13259,vllm13084,vllm13610,vllm15330,vllm9722}, particularly under dynamic workloads with mixed task types and fluctuating request rates.





\subsection{Related Work}
\label{sec:background:related}
\subsubsection*{Batch Scheduling in Traditional DNN Serving}
Prior work has explored dynamic batching to meet SLA targets in traditional DNN serving pipelines.
Zhang et al.~\cite{zhang2023shepherd}, Gujarati et al.~\cite{gujarati2020serving}, and Romero et al.~\cite{romero2021infaas, romero2021llama} proposed dynamic batch size tuning techniques motivated by the tradeoff between queueing delays and execution latency as batch sizes vary.

These methods typically rely on offline profiling to collect execution times at different batch sizes and use these latency tables during inference to guide online batching decisions. For example, Clockwork~\cite{gujarati2020serving} uses precise timing predictions to orchestrate synchronized batched execution across requests.

However, these approaches assume fixed and predictable latency profiles, which are unsuitable for decoder-based LLMs where latency is input- and context-dependent due to auto-regressive generation. In contrast, our work targets continuous batching for LLMs and explicitly handles uncertainty in token-level compute dynamics.

\subsubsection*{Request Interference in Continuous Batching} 
Recent studies have analyzed the impact of request interference in continuous batching. 

Sun et al.~\cite{sun2024llumnix} quantify this interference, showing that decode speed degrades significantly as total tokens per batch grow, with up to 2.6× variation depending on batch size and composition. To mitigate this, they propose dynamic request migration across GPUs.

Other works~\cite{zhong2024distserve, agrawal2023sarathi, hu2024inference, patel2024splitwise} further separate prefill and decode phases onto different hardware to reduce cross-phase interference. While these strategies are effective in multi-GPU or datacenter-scale deployments, they do not address the more common and practical settings for small development teams or indie developers: serving models on single-GPU or resource-constrained environments, which is our primary focus.

\subsubsection*{SLA-Aware Scheduling}
Some systems address latency sensitivity by aggressively reordering requests. ExeGPT~\cite{exegpt} and Niyama~\cite{niyama} propose SLA-aware scheduling algorithms that reorder requests based on latency targets. However, these approaches require substantial modifications to the inference engine internals and have not yet been integrated into widely-used serving stacks, which is not easy for small team or indie developers to begin with.

Our approach departs from above work by offering a lightweight, non-intrusive mechanism that adapts to request heterogeneity and load variation in real time, making it directly applicable to production-grade LLM serving engines like vLLM.


%% file: sections/3.motivation.tex
\section{Motivation: Static Configurations for Continuous Batching Are Not Always Optimal}
\label{sec:motivation}

This section empirically shows that a static batch size cannot provide optimal service‑level compliance when either the task mix or the request load shifts, a common occurrence in interactive development workflows. We first outline the hypothesis, then describe the experimental setup, and finally extract two key observations that motivate the need for an adaptive batching mechanism.

\subsection{Hypothesis}
The static batch size controls the largest batch size that can be processed at once. We hypothesize that static configurations face two fundamental limitations:

\begin{itemize}
    \item \textbf{Workload Composition Sensitivity} A single static configuration cannot deliver optimal goodput across heterogeneous workload mixes as different task types contend for compute resources.
    \item \textbf{Request Load Sensitivity} Even for a fixed workload composition, the optimal configuration shifts as request rate (RPS) varies.
\end{itemize}

\subsection{Experimental Setup}

Our experiments simulate a resource-constrained setting where a CodeLLM-based (Qwen-Coder-2.5B model~\cite{hui2024qwen2}) coding assistant is served on a single  accelerator unit using \texttt{vLLM}~\cite{vllm}.


We use four representative curated datasets covering common coding tasks listed in Table~\ref{tab:task_types} along with the sources, average input and output tokens, and simulated SLAs:

\begin{itemize}
\item \textbf{Code QnA}~\cite{hl-codellama-chat-response}: Interactive coding-related questions and answers, representing quick developer queries with short answers.
\item \textbf{Code Generation}~\cite{synthetic-code-generations}: Natural language descriptions paired with generated code implementations.
\item \textbf{Code Summary}~\cite{code-summary-java}: Short code segments paired with summaries, simulating documentation tasks.
\item \textbf{Code Translation}~\cite{code-translation}: Source code in one language paired with equivalent code in another language.
\end{itemize}


\begin{table}[h]
\centering
\begin{tabular}{|l|c|c|c|}
\hline
\textbf{Task Type} & \makecell{\textbf{Average}\\\textbf{Input Tokens}} & \makecell{\textbf{Average}\\\textbf{Output Tokens}} & \textbf{SLA (s)} \\ \hline

Code QnA~\cite{hl-codellama-chat-response}    & 186    & 43    & 1   \\ \hline
Code Generation~\cite{synthetic-code-generations}   & 463  & 387   & 8     \\ \hline
Code Summary~\cite{code-summary-java}    & 31  & 30 & 1  \\ \hline
Code Translation\cite{code-translation}        & 670     & 617  & 12  \\ \hline
\end{tabular}
\caption{Task types used in our workload.}
\label{tab:task_types}
\end{table}

To simulate the Service Level Agreements (SLAs), we use the mean completion time when serving 100 requests at 10 RPS (considered as moderate load in our setting) with a Poisson arrival distribution.
We then construct three representative workloads with different mixes of the tasks for evaluation under different usage scenarios listed below:


\begin{itemize}
  \item \textbf{W1 (Heavy)}: majority of the tasks generate longer tokens, 40 \% Code Translation, 40 \% Code Generation, 10 \% Code QnA, and 10 \% Code Summary.
  \item \textbf{W2 (Light)}: majority of the tasks generate shorter tokens, 40 \% Code QnA, 40 \% Code Summary, 10 \% Code Generation, and 10 \% Code Translation.
  \item \textbf{W3 (Balanced)}: 25 \% of each task for a balanced mix of tasks.
\end{itemize}

To evaluate whether static configurations of batch size can meet the performance demands of 3 real-world workloads (defined above), we evaluate 10 batch size configurations from 10 to 100 in increments of 10, as values above 100 exceeded the GPU memory capacity on our hardware. We experiment with 12 request rates (1–10, 15, 20 RPS), yielding $3 \times 12 \times 10 = 360$ configurations. We select RPS values of 1–10 to observe gradual performance changes, and include higher rates of 15 and 20 RPS to examine system behavior under heavy load conditions.
For each configuration, we send 100 requests following a Poisson distribution, which simulates real usage patterns~\cite{jha2024learnedbesteffortllmserving, srivatsa2024prebleefficientdistributedprompt}. 

Each configuration is repeated three times and we only report the average metrics across three runs.

We use \textbf{Goodput} as our primary metric defined as the percentage of requests that complete successfully within their respective SLA thresholds defined in Table~\ref{tab:task_types}: 
\[
\texttt{Goodput } = \frac{\texttt{\# of requests completed within SLA}}{\texttt{Total number of requests}}
\]
 A request is only considered successful if its end-to-end latency is less than or equal to the SLA for its task type. This metric reflects how well a system meets practical developer expectations under real-world load~\cite{gong2025past, zhong2024distserve}.

\subsection{Observations}

Our experiments uncover two key limitations of static batch sizes.

\paragraph{Observation 1: No single best configuration.}
Figure~\ref{fig:rq1_results} shows that the best setting varies widely across workloads: 30 for W1, 70 for W2, and 80 for W3. If adopting the optimal configuration from one workload to another yields disastrous results: using W2's optimal setting (70) for W1 produces extremely low goodput at most RPS levels. Even small changes in configuration (e.g., from 30 to 50 in W3) can cause goodput to drop sharply( up to 39\%) due to interference between task types. This confirms our hypothesis that workload composition heterogeneity makes it hard to pick a single static setting that works well across.

\paragraph{Observation 2: Optimal configuration changes with request load}
The blue boxes in each heatmap, marking the optimal configuration for each RPS level. Critically, these optimal configurations vary significantly across different request load rates. For instance, in Workload 3, the optimal batch size is 100 at RPS is 5, but shifts to 80 at RPS is 10, and further changes in higher load. This demonstrate that no single static configuration can remain optimal as request intensity fluctuates. This variability forces developers to choose between frequent system restarts to adjust configurations or accepting significant performance penalties. Such static configurations are thus unsuitable for production environments with dynamic workloads.

%% file: sections/4.system.tex
\begin{algorithm}[ht]
\caption{Admission Control Process}
\label{alg:coordinator-decision}
\begin{algorithmic}[1]
\State HPQ $\gets$ High Priority Queue
\State LPQ $\gets$ Low Priority Queue
\State \textbf{function} AdmissionControl():
\While{true}
    \If{not IsEmpty(HPQ)}
        \State reqWin $\gets$ RandomSample(HPQ, windowSize)
        \For{req in reqWin}
            \State predSpd $\gets f(\text{curLoad} + 1)$
            \State reqSpd $\gets$ req.maxTokens / (req.deadline - currentTime)
            \If{predSpd $<$ reqSpd $\vee$ predSpd $<$ any(actReq.reqSpd)}
                \State continue
            \Else
                \State AdmitToBatch(req)
                \State \textbf{break}
            \EndIf
        \EndFor
    \Else
        \If{not IsEmpty(LPQ)}
            \State req $\gets$ Dequeue(LPQ)
            \State Execute(req)
        \EndIf
    \EndIf
    \State Sleep($\delta$)
\EndWhile
\end{algorithmic}
\end{algorithm}

\section{Our Approach}
\label{sec:approach}

This section presents~\textsc{SABER}, our approach towards SLA-aware continuous batching. Section~\ref{sec:approach:overview} provides an overview of the approach. Section~\ref{sec:profiling} details the Offline profiling phase. Section~\ref{sec:approach:online} describes the online serving phase and its admission control mechanism.

\subsection{Overview}
\label{sec:approach:overview}

\begin{figure}[t]
    \centering
    \includegraphics[width=1\linewidth]{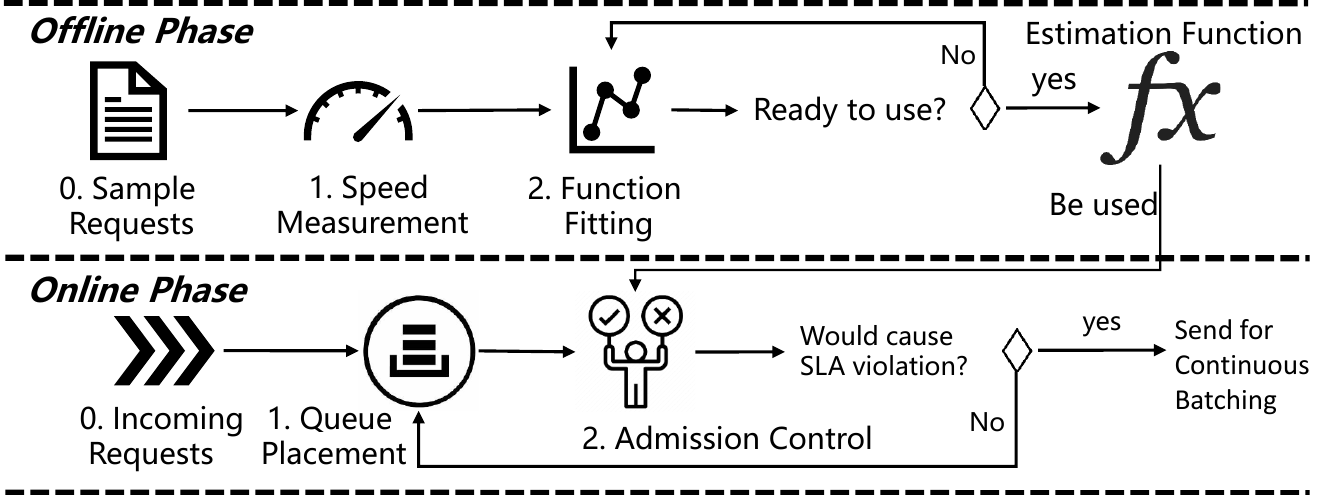}
\caption{The \textsc{SABER} workflow. \textit{Offline}: profiles workloads and trains estimation function $f(L)$ for token generation speed prediction. \textit{Online}: uses $f(L)$ to admit requests that won't violate existing SLAs, deferring others to maintain system performance.}
    \label{fig:overview}
\end{figure}

Figure~\ref{fig:overview} illustrates the two-stage workflow of our approach: the Offline phase and the Online phase. In the Offline phase, we first profile diverse workloads. Based on the profiling results, we fit an estimation function that can predict the causal impact of admitting incoming requests into the batch queue on performance. This function supports the admission control mechanism in the online phase.

During Online phase, incoming model requests are processed via three steps. First, requests enter a two-tier queue that continuously tracks each request's \texttt{remaining time} to SLA. Second, an admission control mechanism uses the trained estimation function to evaluate whether admitting a new request would make any requests in execution violate their SLA constraint. Only requests that are admitted can proceed to the third step, i.e., joining the batch queue.

\subsection{Offline Phase}
\label{sec:profiling}
This section describes the Offline phase of \textsc{SABER}. 
The main objective of this phase is to train an estimation function that can predict generation speed under concurrent request execution. For CodeLLM serving, estimating end-to-end latency directly is challenging due to nondeterministic number of tokens generated by decoder-only transformer-based language models. 

Instead of estimating the end-to-end latency of requests, We observe that focusing on generation speed (tokens per second) provides a more reliable result, as the execution time can be derived by dividing the required output tokens by the generation speed. In addition, for certain types of workload, we can also estimate the average generated tokens~\cite{sharegpt_dataset2023}.
Hence, by collecting empirical performance data to fit an estimation function capable of predicting how additional requests affect generation speed, we can approximate the end-to-end latency of a request. The whole phase shown in Figure \ref{fig:overview} consists of two sequential steps:


\emph{Step 1: Token Generation Speed Measurement.} Using the workload samples prepared in Step 0, in this step, we configure the system with a large batch size value that exceeds the maximum number of concurrent requests we would send, then continuously inject mixed workloads of varying sizes. We monitor each request's execution time and number of generated tokens under different system loads. This process produces a training data consisting of load-speed pairs: for each request, we record the system load $L$  (number of concurrent requests) during its execution and calculate its token generation speed  as \texttt{generated tokens} / \texttt{execution time}.

\emph{Step 2: Estimation Function Fitting.} In this step, we train an estimation function based on the performance measurements collected in the previous step. The estimation function
\[
\hat{v} = f(L)
\]
predicts token generation speed, where \( L \) denotes the current number of executing requests. In the training process, we utilize approximately 1,000 data points collected across different concurrency levels and workload compositions. We evaluate several modeling approaches and select the Universal Scalability Law (USL)~\cite{wang2018integrating} with the best R\textsuperscript{2} ~\cite{rousson2007r}, using \texttt{scipy.optimize} library~\cite{virtanen2020scipy} for curve fitting.

In summary, the output of the offline phase is an \emph{estimation function} that approximates the token generation speed by considering current active requests in the batch queue of the CodeLLM.
For example, given a running CodeLLM instance with 5 active requests and an incoming request, the function predicts, when the requests are admitted, on what extent the per-request token generation speed may drop. The estimation function plays a key role in the Online Phase to decide whether a request can be allowed to join continuous batching execution with SLA constraint. The details are described next in Section \ref{sec:approach:online}

\subsection{Online Phase}
\label{sec:approach:online}

During the Online phase, incoming requests flow through the pipeline 
illustrated in Figure~\ref{fig:overview}. While the complete pipeline 
includes request arrival (Step 0) we focus on the two core steps that embody \textsc{SABER}'s key innovations:  queue placement (Step 1) and admission control (Step 2).  \textsc{SABER}'s core decisions occur in Steps 1-2, where requests are placed into a two-tier queue and evaluated for admission based on SLA constraints. We detail these core mechanisms below:

\emph{Step 1: Queue Placement}
The queue employs a two-tier structure to prioritize feasible requests. All incoming requests initially enter the high-priority queue with two key attributes:
\begin{enumerate}
    \item \textbf{Deadline (Tail-Latency SLA):} The completion deadline for this request defined by the user or system administrator.
    \item \textbf{\texttt{Max Number of Generated Tokens}:} The maximum number of tokens the model expected to generate for this request.
\end{enumerate}

\[
\texttt{required speed} = \frac{\texttt{Max \# of Generated Tokens}}{\texttt{remaining time}}
\]
As requests wait in the queue, their \texttt{remaining time} (calculated as the difference between the deadline and current time) decreases and thus the \texttt{required speed} increases: For instance, a request that is required to finish generating 500 tokens estimatingly within 10 seconds slack remaining needs the \texttt{required speed} to be 50 tokens/second. After waiting in the queue for 5 seconds, the request now needs the \texttt{required speed} to be 100 tokens/second.

We monitor the speed requirements against the model serving engine's capability. When a request's \texttt{required speed} exceeds model serving engine's fastest speed (e.g 100 tokens/s when only serve one request but request requires 120 tokens/second to meet SLA), it becomes infeasible for this request to ever meet its SLA. Such requests are automatically demoted to the low-priority queue, which is processed only when the high-priority queue is empty.
\begin{figure*}[t] 
    \centering
    \begin{subfigure}[t]{0.33\textwidth} 
        \includegraphics[width=\textwidth]{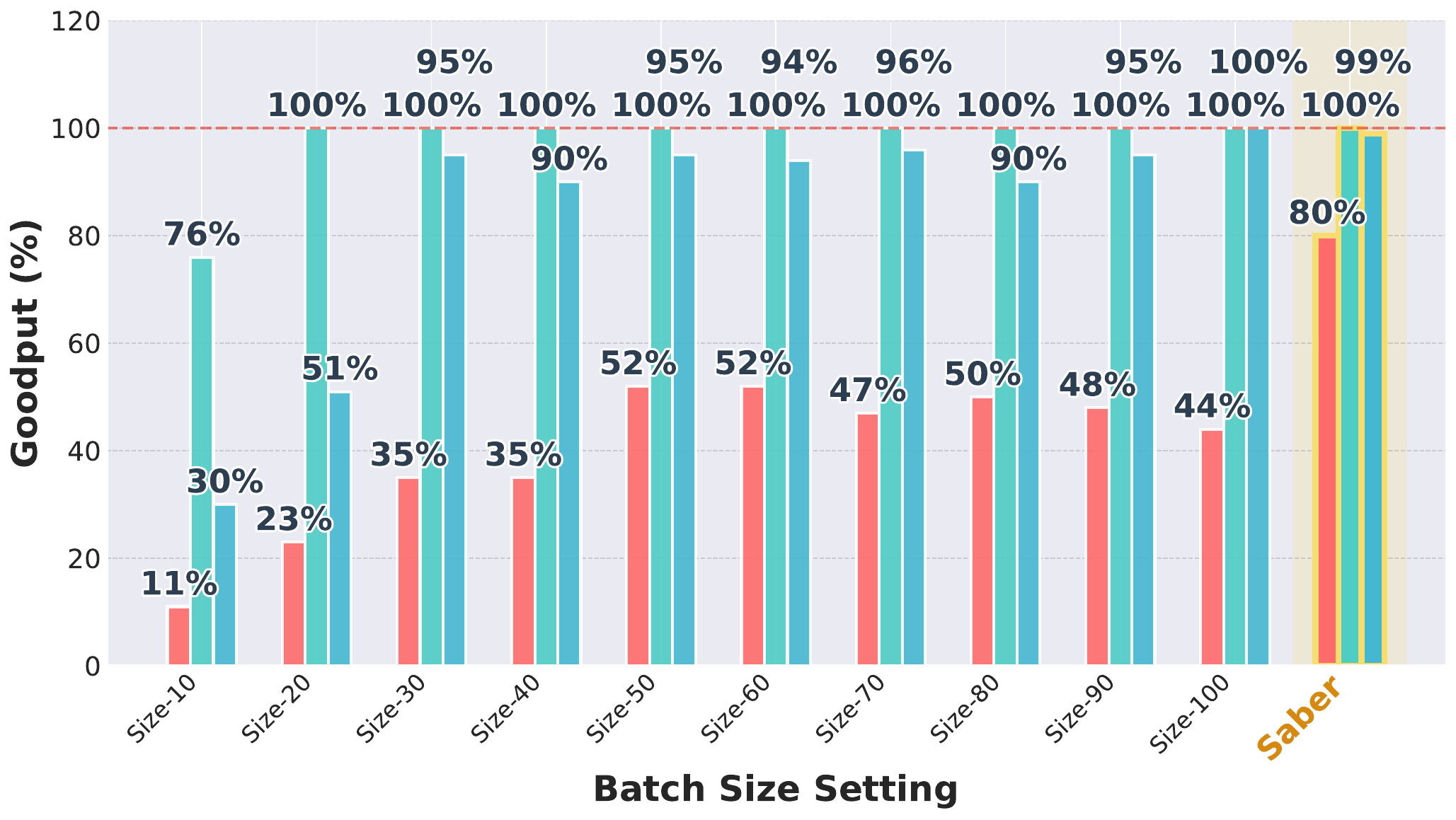}
        \caption{Low load (5 RPS)}
        \label{fig:rq2_rps_5}
    \end{subfigure}%
    \begin{subfigure}[t]{0.33\textwidth} 
        \includegraphics[width=\textwidth]{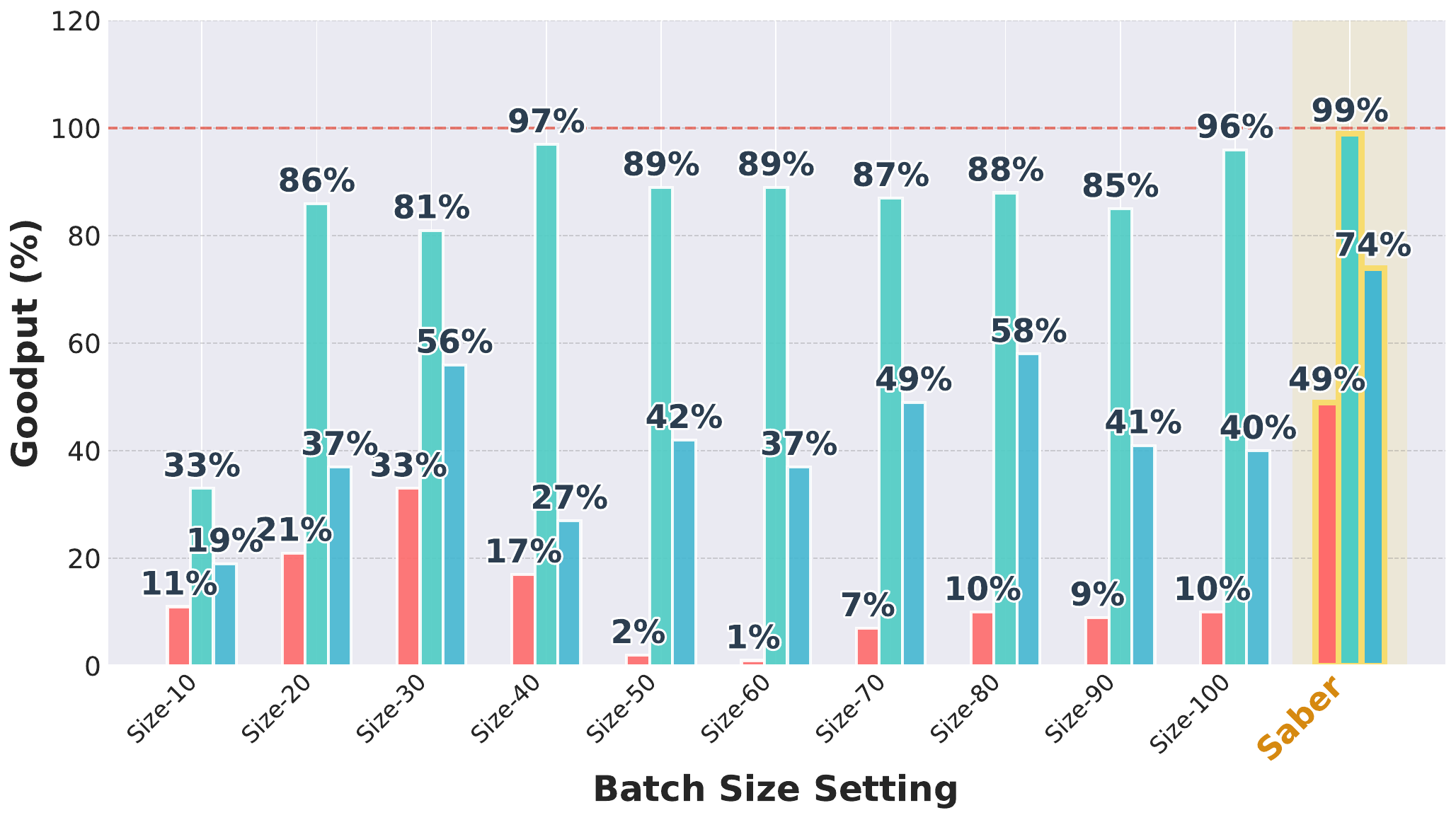}
        \caption{Moderate load (10 RPS)}
        \label{fig:rq2_rps_10}
    \end{subfigure}%
    \begin{subfigure}[t]{0.33\textwidth} 
        \includegraphics[width=\textwidth]{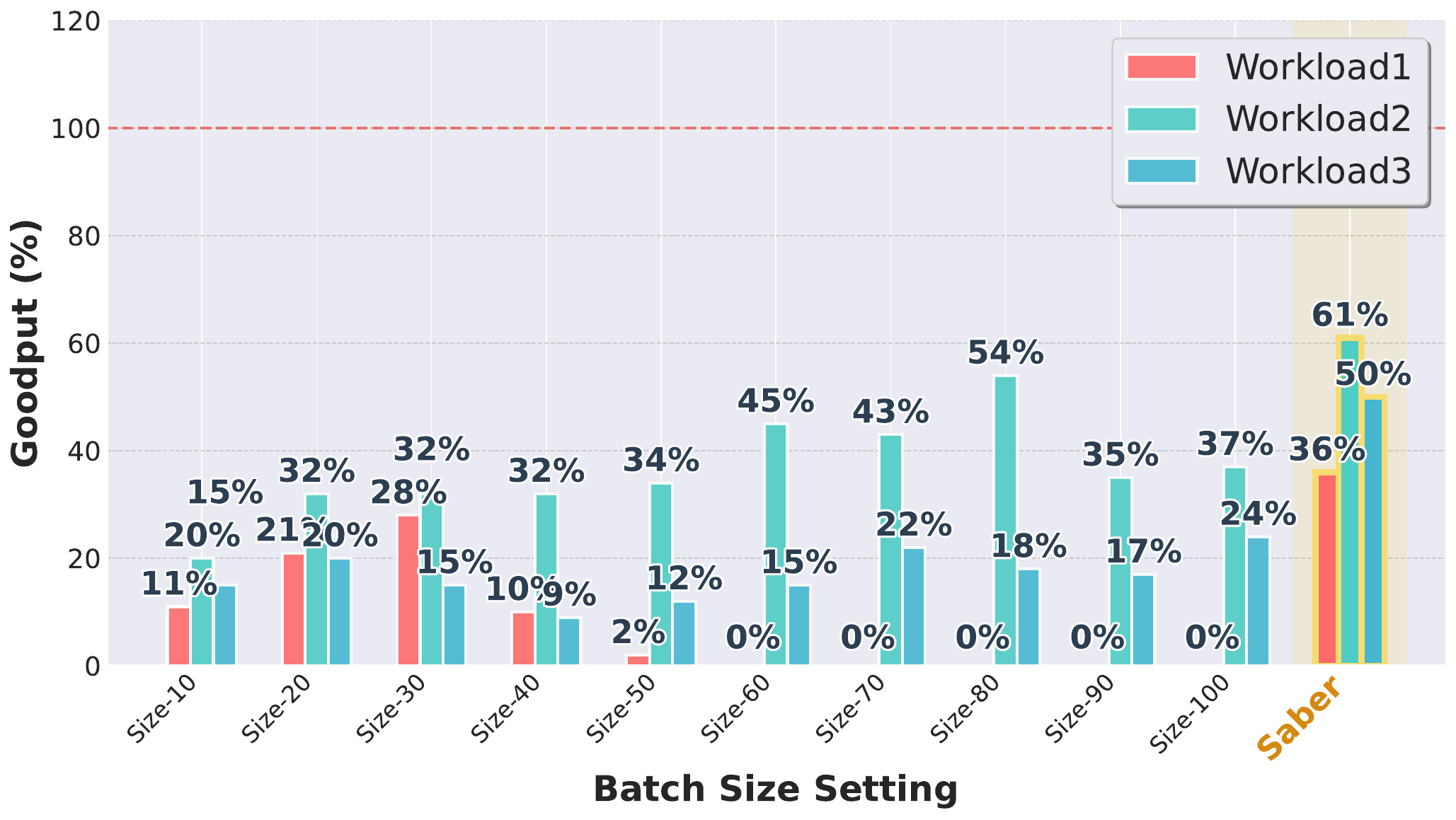}
        \caption{High load (20 RPS)}
        \label{fig:rq2_rps_20}
    \end{subfigure}
    \caption{Goodput comparison between static concurrency configurations and Saber across different request rates. Each subfigure shows the percentage of requests meeting SLA requirements for three workload types under varying concurrency limits (10-100). Saber dynamically adapts its concurrency to maintain optimal performance across all load conditions.}
    \label{fig:RQ2}
\end{figure*}
\emph{Step 2: Admission Control}
This step evaluates whether to admit the selected request into the batch queue. As shown in Algorithm ~\ref{alg:coordinator-decision}, the admission control runs continuously (line 4), and processes requests from the high-priority queue when available (line 5). The mechanism first randomly selects one request from first \texttt{n} requests in the batch queue head (where \texttt{n} is the \texttt{windowSize}) to avoid head of line blocking. For the selected request, it performs a two-fold evaluation: First, it uses the estimation function with the current load plus one to predict the per-request generation speed if the selected incoming request were admitted to the batch execution (line 8). Second, it checks two conditions (line 10): 1) Whether the predicted speed meets the incoming request's \texttt{required speed}. 2) Whether admitting this request would cause any currently executing request to violate its SLA. If either condition fails, the request is NOT admitted and remains in the high priority queue for future evaluation. Otherwise, it is admitted to the batch execution (line 13), as shown as request inference in the Figure \ref{fig:overview}. \textsc{SABER} also records this request's required generation speed to meet SLA with the current timestamp. 

When the high-priority queue is empty (line 18), the system processes low-priority requests without speed estimation (lines 18-20), as these requests have already missed their SLA targets and are served on a best-effort basis.

%% file: sections/5.RQ1.tex
\begin{figure*}[t]
   \centering
   \includegraphics[width=0.95\linewidth]{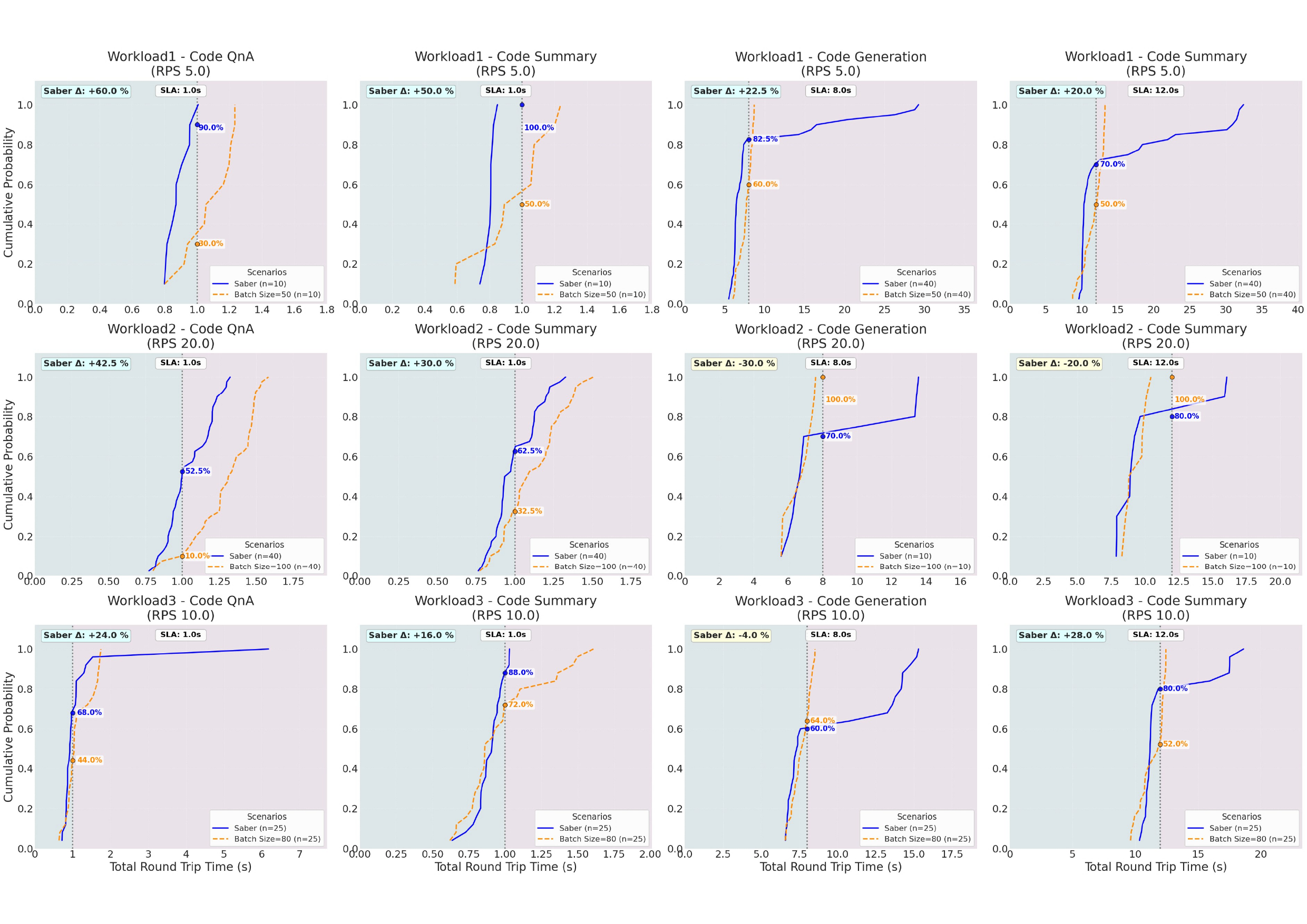}
   \caption{Cumulative Distribution Function (CDF) analysis of request completion times at RPS=10. Each subplot shows the percentage of requests (y-axis) completing within a given time (x-axis) for different task types. Vertical dashed lines indicate task-specific SLA thresholds. For fair comparison, we use the optimal static configuration for each workload (e.g., max\_concurrency=60 for Workload1).}
   \label{fig:CDF}
\end{figure*}

\section{RQ1: How well does \textsc{SABER} sustain SLA compliance across varied workload compositions?}
\label{sec:rq1}

In this section, we evaluate \textsc{SABER}'s ability to maintain optimal goodput compared to static configurations under diverse workload compositions. Section~\ref{sec:rq1:motivation} describes the motivation behind this RQ. Section~\ref{sec:rq1:exp} explains the experiment setup. Section~\ref{sec:rq1:results} presents the findings and implications.

\subsection{Motivation}
\label{sec:rq1:motivation}

The results of our motivational experiments in Section~\ref{sec:motivation} reveal a limitation in existing LLM serving engines that support continuous batching, i.e., a batch size configuration that yields optimal goodput for one workload composition (e.g., 30 for Workload 1) can perform suboptimally for another (e.g., 30 for Workload 2). The sensitivity to workload composition underscores the necessity of an adaptive approach to achieving optimal goodput for evolving workload mixes. This RQ aims to evaluate whether~\textsc{SABER} is able to outperform the static configurations across three representative workload compositions.

\subsection{Experiment Setup}
\label{sec:rq1:exp}

In this experiment, we keep the same setup as Section \ref{sec:motivation} with a single accelerator unit as the hardware environment and three types of workload compositions for RPS 5, 10, and 20. We serve the same CodeLLM (Qwen-Coder-2.5B) for the experiment with vLLM.

We reuse the experiment results in the motivation example conducted in Section~\ref{sec:motivation}, where we collected the goodput for each static configuration of \textsf{batch size} under the RPS of 5, 10, and 20 as they represent light, medium, and heavy loads. Then we ran the same workloads shown in Table~\ref{tab:task_types} with~\textsc{SABER} under the same RPS. 

We first compare the goodput between~\textsc{SABER} and the static configurations under each RPS setting. In addition to reporting goodput, we analyze the distribution of request completion times using Cumulative Distribution Function (CDF) plots~\cite{6008726, 5581588}. The CDF is a standard tool for characterizing system performance, as it captures the entire distribution—including median, percentile, and tail behaviors—essential for evaluating SLA compliance.

\subsection{Results}
\label{sec:rq1:results}

\textbf{~\textsc{SABER} consistently outperforms static configurations across diverse workload compositions.} Figure~\ref{fig:RQ2} presents a comparison of goodput between \textsc{SABER} and static batch size configurations, evaluated across three distinct workload compositions and three representative request rates (RPS). Specifically, we select RPS values of 5, 10, and 20 to illustrate system behavior under normal (RPS=5), moderate (RPS=10), and heavy (RPS=20) load conditions. We omit lower RPS values (RPS less than 5) from the analysis, as minimal resource contention in this regime allows most configurations to achieve near-perfect goodput, providing little insight into comparative performance.

Across all evaluated settings, \textsc{SABER} consistently outperforms static batch size configurations, achieving higher goodput under every workload composition and load level (see Figure~\ref{fig:RQ2}). Below, we detail the specific improvements observed in each scenario:

\begin{itemize}
\item \textbf{W1 (Heavy)}: Under all RPS levels, static configurations suffer significantly—especially at high load, where large batch sizes can lead to severe resource contention and, in extreme cases, 0\% goodput. In contrast, \textsc{SABER}'s adaptive strategy ensures critical requests are prioritized and completed before SLA deadlines (36\% goodput, 8\% higher than the most optimal static configurations),

\item \textbf{W2 (Light)}: Here, most requests quickly complete the prefill phase, allowing for greater parallelization. While static configurations fare better than in workload 1, \textsc{SABER} still maintains an advantage. At RPS 20, \textsc{SABER} achieves 63\% goodput, outperforming the best static configuration by 7\%.

\item \textbf{W3 (Balanced)}: The mixed scenario is especially challenging for static settings, which must compromise between competing requirements. \textsc{SABER} dynamically adapts to the shifting task mix, resulting in consistently superior performance. At RPS 10, \textsc{SABER} achieves 74\% goodput versus 56\% for the best static configuration, an 18\% improvement.
\end{itemize}

 \textbf{Our approach achieves better performance by de-prioritizing requests that are no longer possible to meet SLAs, reallocating compute to requests that are more likely to meet SLAs to avoid contention.} We now explore the rationale through detailed CDF analysis. For each workload, we select the load level where SABER has the most improvements and compare it against the best-performing static configuration: Workload 1 at RPS=5 with static \textsf{batch size} 50, Workload 2 at RPS=20 with \textsf{batch size} 100, and Workload 3 at RPS=10 with \textsf{batch size} 80. For each workload, we decompose the results into task level CDF for each task type. The results are shown in Figure~\ref{fig:CDF}. Each row fixes a workload composition and request rate (RPS); each column fixes one of the four task types. The x-axis represents completion time in seconds and the y-axis shows the cumulative percentage of completed requests (0-100\%). Vertical dashed lines mark task-specific SLA thresholds from Table~\ref{tab:task_types}. Hence, the green area to the left of the dashed SLA line marks requests that finish in time; the red area on the right represents SLA violations.

\begin{figure*}[ht]
    \centering
    \begin{subfigure}[t]{0.33\textwidth} 
        \includegraphics[width=\textwidth]{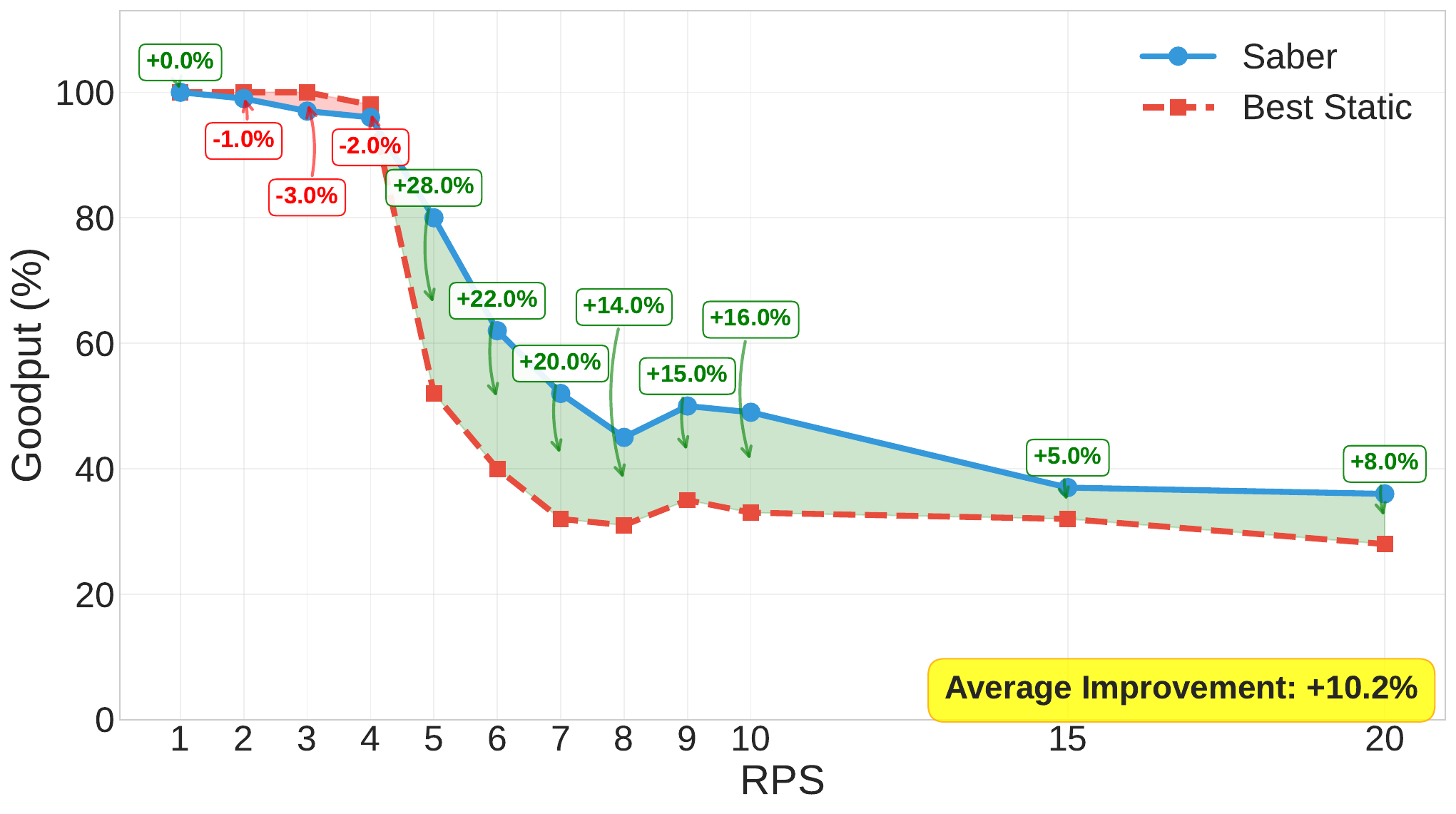}
        \caption{Workload 1}
        \label{fig:goodput_workload1}
    \end{subfigure}%
    \hfill
    \begin{subfigure}[t]{0.33\textwidth} 
        \includegraphics[width=\textwidth]{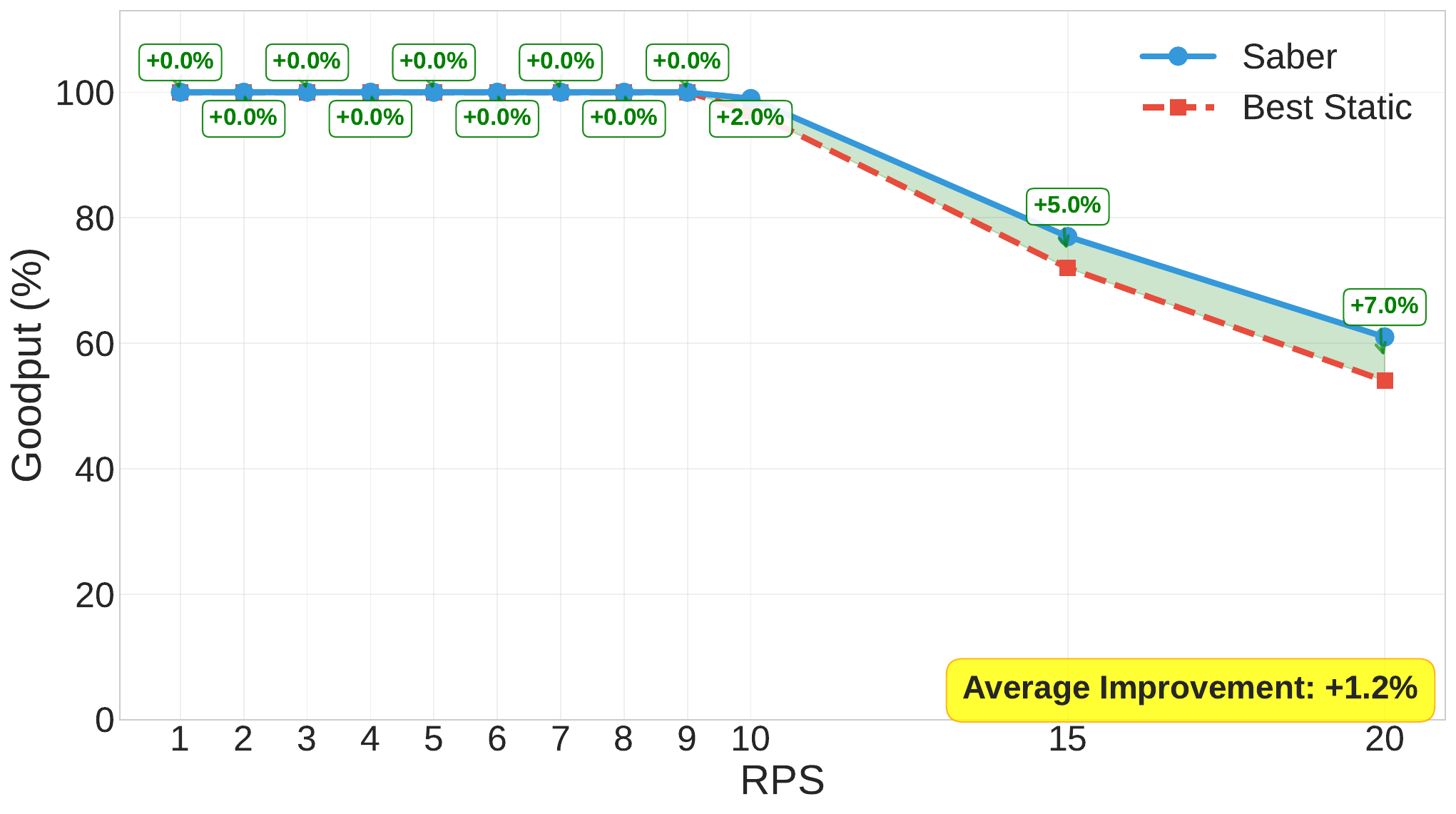}
        \caption{Workload 2}
        \label{fig:goodput_workload2}
    \end{subfigure}%
    \hfill
    \begin{subfigure}[t]{0.33\textwidth} 
        \includegraphics[width=\textwidth]{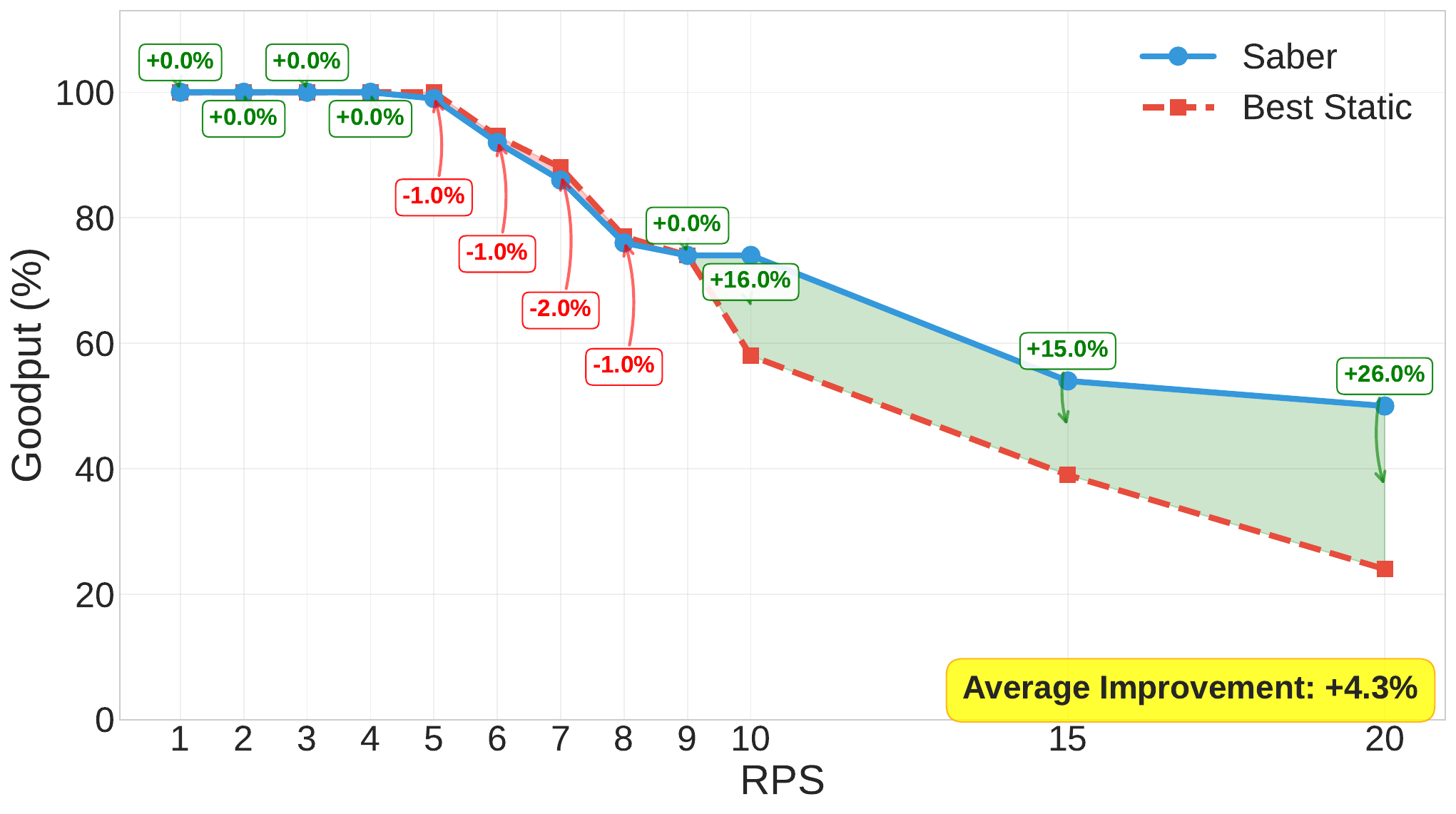}
        \caption{Workload 3}
        \label{fig:goodput_workload3}
    \end{subfigure}%
    \caption{Goodput comparison between \textsc{SABER} and best static configurations across varying loads.}
    \label{fig:RQ3_goodput}
\end{figure*}
\begin{itemize}
\item \textbf{W1 (Heavy) RPS = 5.}
Load is moderate, so \textsc{SABER} can improve every task simultaneously.
The blue curves rise much more steeply before the SLA line(e.g.\ CodeQnA +60\%.\ and CodeSummary +50\%), while the residual tail beyond 35 s shows that a handful of oversized requests are deliberately postponed. 

\item \textbf{W2 (Light)  RPS = 20.}
At high load, \textsc{SABER} deliberately triages the queue: it accelerates light tasks (code QnA +42.5\%., code summary +30\%.) while accepting a drop for heavy code generation and code translation ( -30\%. and -20\%., respectively). This policy not only prevents head-of-line blocking, evidenced by the near-zero goodputs of the static baseline, but also matches user-experience priorities: studies of interactive services show that users are far less tolerant of delays on tasks they expect to complete quickly, whereas they are willing to wait longer for inherently heavyweight operations. By reallocating GPU time accordingly, \textsc{SABER} maximises perceived responsiveness where it matters most while containing the overall SLA violation rate.

\item \textbf{W3 (Balanced), RPS=10.}
With a heterogeneous mix and moderate load level, \textsc{SABER} adopts a middle ground.  Light tasks still gain markedly (code QnA +24\%., code summary +16\%); heavy code generation is roughly on par ( -4\%.) while code translation jumps +28\%.
The tails are shorter than in workload 2, showing that adaptive batching now has enough headroom to rescue a portion of heavy requests without starving the light ones.
\end{itemize}

Overall, the per-task CDFs reveal the mechanism behind the aggregate goodput gains shown in Figure~\ref{fig:CDF}: \textsc{SABER} continuously predicts each request’s likelihood of meeting its SLA and reassign tasks aggressively under heavy pressure, more proportionally when load is lighter, thereby maximising user-visible success while containing the long-tail latency of inevitable stragglers.

\begin{tcolorbox}[colback=white, colframe=black, boxrule=1pt, arc=0pt, left=10pt, right=10pt, top=10pt, bottom=10pt]
\textbf{Findings:} (1) \textsc{SABER} consistently outperforms the best static batch configurations by 1.2 to 10.2\% across diverse workload compositions without requiring manual tuning. (2) These performance improvements stem from proactively identifying requests unlikely to meet SLAs and strategically deprioritizing them, thereby reallocating GPU resources to requests with higher completion probabilities. This behavior is evidenced by the distinctive long-tail distributions in request completion time CDFs.

\textbf{Implications:} Static batch size configurations struggle under heterogeneous and dynamic workloads. Effective LLM serving systems require adaptive, SLA-aware batching strategies capable of real-time prioritization, especially in resource-constrained scenarios. Prioritizing requests according to user-perceived responsiveness ensures higher overall system efficiency and improved user experience.
\end{tcolorbox}

%% file: sections/6.RQ2.tex
\section{RQ2: To what extent can \textsc{SABER} dynamically adapt to unpredictable request loads without manual intervention?}
\label{sec:rq2}

In this section, we evaluate \textsc{SABER}'s performance under different levels of loads, in comparison with the performance of the best static configurations under the same level of loads. Section~\ref{sec:rq2:motivation} describes the motivation behind this RQ. Section~\ref{sec:rq2:exp} explains the experiment setup. Section~\ref{sec:rq2:results} presents the findings and implications.

\subsection{Motivation}
\label{sec:rq2:motivation}



The results of RQ1 establish that \textsc{SABER} outperforms all static configurations across multiple workload compositions, under representative request rates. However, in production deployments, workload intensity itself is highly dynamic. For example, an internal LLM service might see 2–3 RPS during team standups, spike to 15–20 RPS during peak development hours, and then subside to 5 RPS during review cycles—representing an order-of-magnitude fluctuation within a single day.

This variability raises two pressing concerns. First, while RQ1 evaluated \textsc{SABER} under a few fixed request rates, it remains unclear whether its performance advantage holds across the full spectrum of load conditions—from light to saturated. Our motivational experiments show that no static batch size configuration can serve all RPS levels effectively, suggesting that dynamic adaptation is crucial for systems facing real-world demand patterns.

Second, beyond peak performance, practitioners care about \emph{predictable degradation}. In high-load conditions, static configurations often exhibit brittle behavior: they may work well under nominal load but collapse when overloaded, producing erratic latencies and violating SLA expectations. For system operators, it is not enough to know that performance is ``better on average" - they must understand how performance behaves under stress, and whether it remains controllable.

\subsection{Experiment Setup}
\label{sec:rq2:exp}

In this experiment, we use the same setup as in Sections~\ref{sec:motivation} and~\ref{sec:rq1}, running on a single accelerator unit. We evaluate \textsc{SABER} using the vLLM serving engine with the Qwen-Coder-2.5B model, under the three workload compositions described in Table~\ref{tab:task_types}.

To assess adaptability to dynamic load conditions, we vary the incoming request rate from 1 to 10 RPS in fine-grained steps and additionally include 15 and 20 RPS to simulate heavy-load scenarios. For each (workload, RPS) pair, we compare \textsc{SABER} against the best-performing static configuration selected from 10 candidate batch sizes defined in Section~\ref{sec:motivation}.

Beyond goodput, we introduce two auxiliary metrics to capture performance stability under varying load:
\begin{figure*}[ht]
    \centering
    \begin{subfigure}[t]{0.33\textwidth} 
        \includegraphics[width=\textwidth]{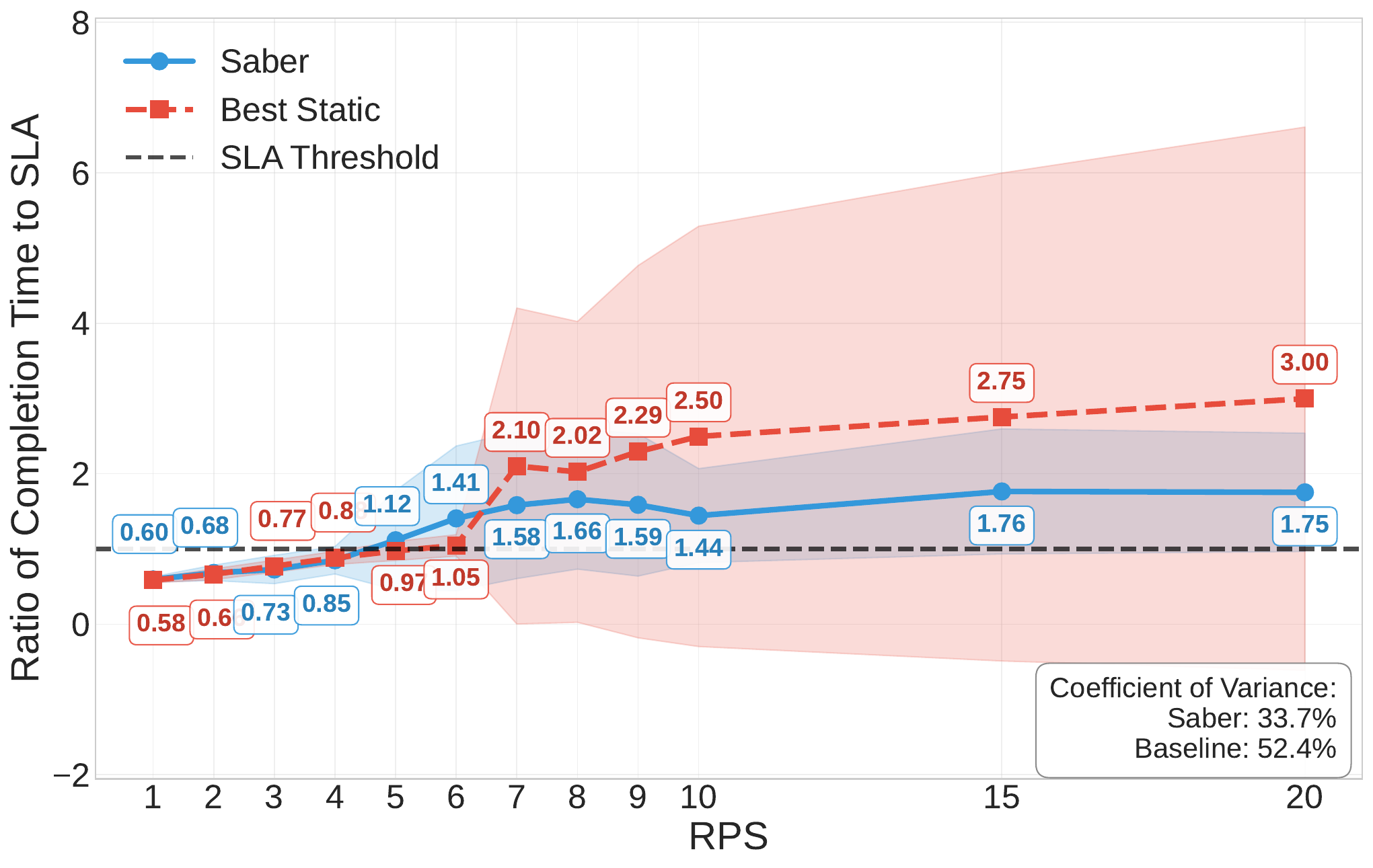}
        \caption{Workload 1}
        \label{fig:stability_workload1}
    \end{subfigure}%
    \hfill
    \begin{subfigure}[t]{0.33\textwidth} 
        \includegraphics[width=\textwidth]
        {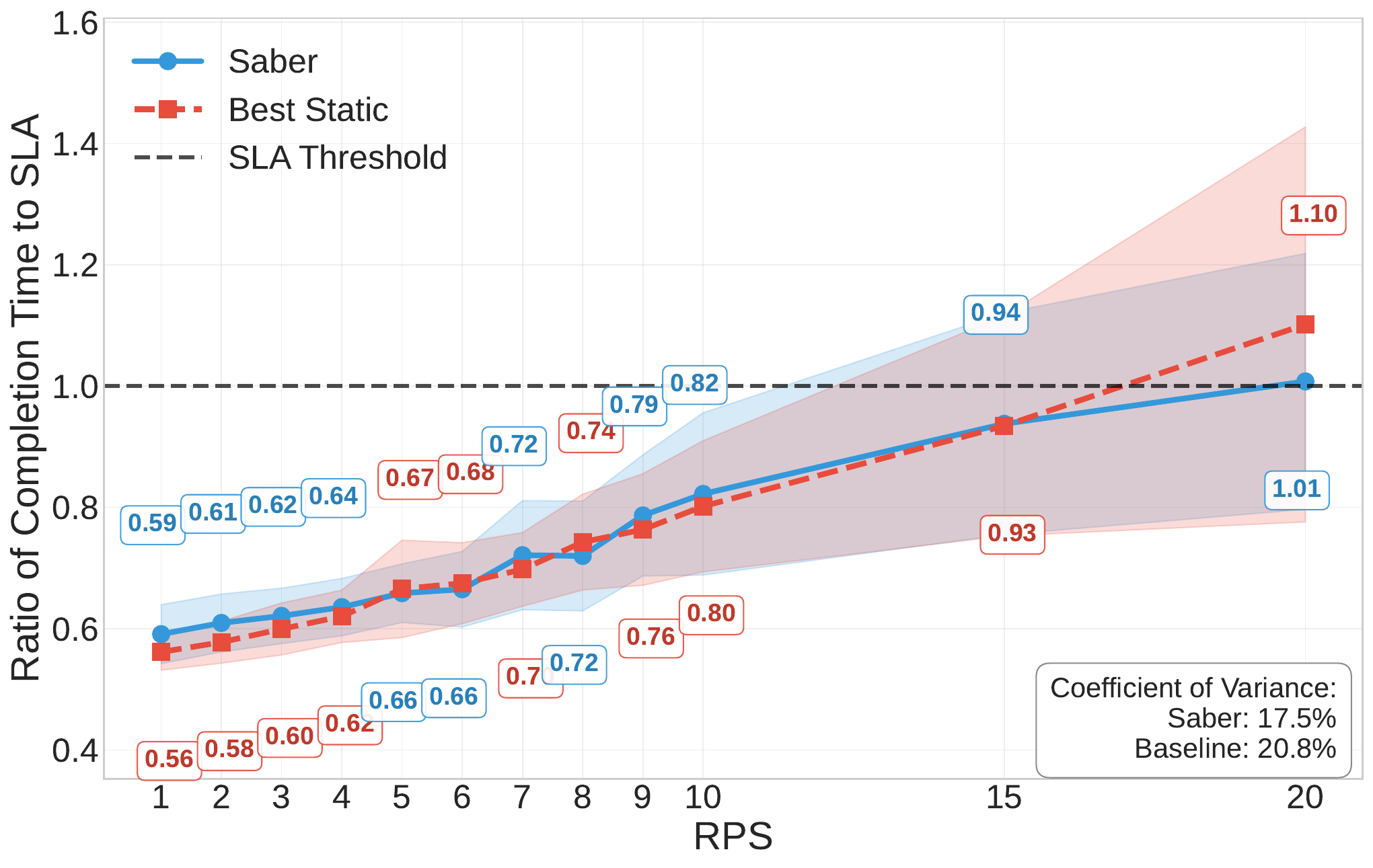}
        \caption{Workload 2}
        \label{fig:stability_workload2}
    \end{subfigure}%
    \hfill
    \begin{subfigure}[t]{0.33\textwidth} 
        \includegraphics[width=\textwidth]
        {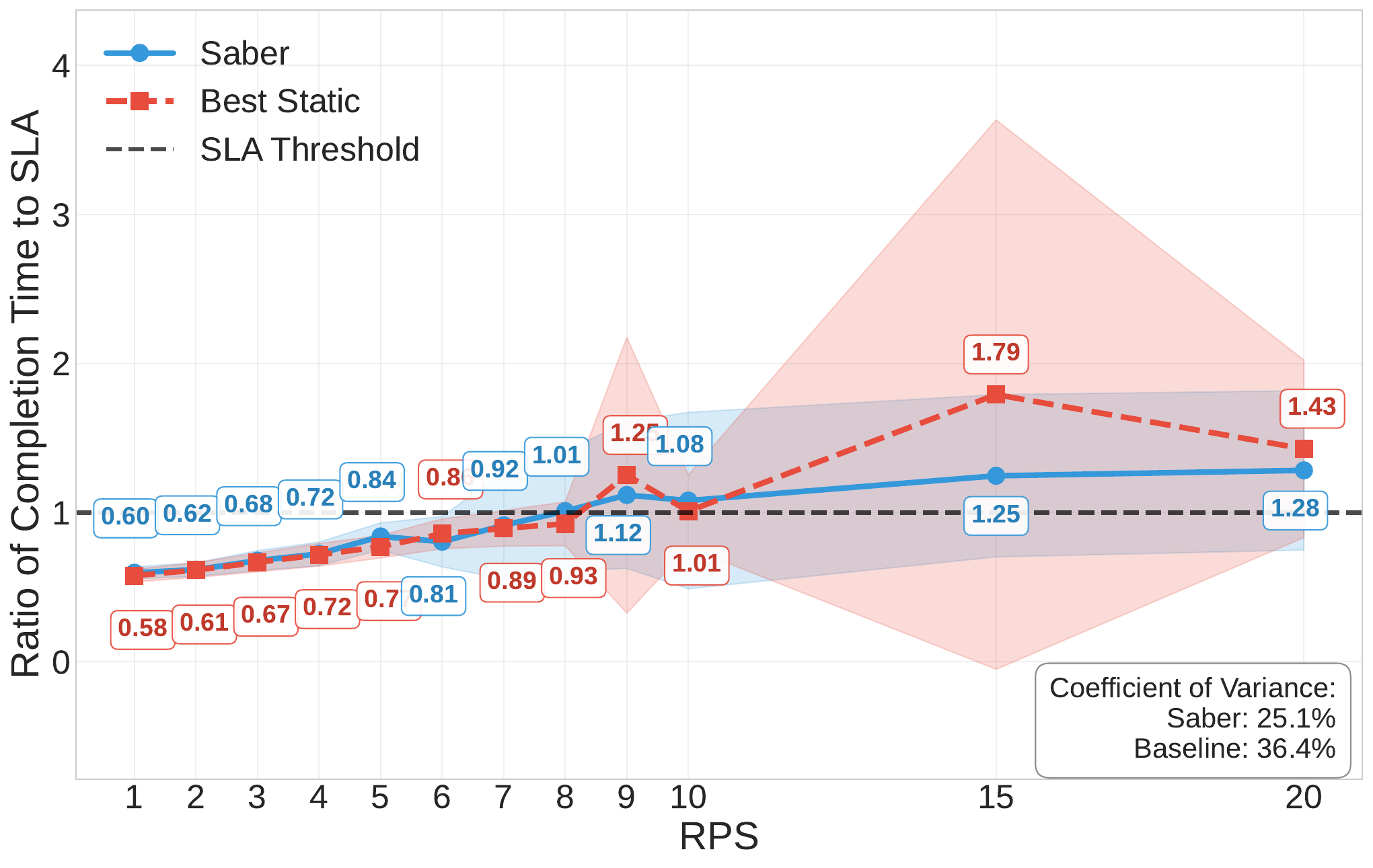}
        \caption{Workload 3}
        \label{fig:stability_workload3}
    \end{subfigure}%
    \caption{Performance stability analysis across varying request loads. Lines show mean completion time normalized by SLA threshold (values below 1.0 indicate success). Shaded regions represent standard deviation, with narrower bands indicating more predictable performance. Text boxes display coefficient of variation (CV) for each approach.}
    \label{fig:RQ3_stability}
\end{figure*}
\begin{itemize}
\item \textbf{Ratio of Completion Time to SLA}: To allow fair comparison across task types with different SLA thresholds (ranging from 1s to 12s),we compute each request's completion time divided by its corresponding SLA. 


\item \textbf{Coefficient of Variation (CV)}: To quantify performance predictability, we calculate the CV of the above ratios across different RPS values:\begin{equation}
CV = \frac{\sigma}{\mu}
\end{equation} where$\sigma$ and $\mu$ are the standard deviation and mean of the completion time to SLA ratios. Lower CV values indicate more stable performance~\cite{wang2012average}.
\end{itemize}

\subsection{Results}
\label{sec:rq2:results}

\textbf{The improvement \textsc{SABER}achieves becomes larger while request loads increases.}
Figure \ref{fig:RQ3_goodput} tracks goodput as the request rate rises. Averaged over the entire load sweep, this translates to +10.2\%, +1.2\%, and +4.3\% of goodput advanatage for \textsc{SABER} for workloads 1–3, respectively. 
When there is no contention (fewer than 4 RPS), both \textsc{SABER} and static configuration achieve near 100\% goodput, and the curves almost overlap. The small dip for \textsc{SABER} comes from its conservative admission control strategy. Once the load rate reaches the level of moderate (5–15 RPS), resource contention cause the best static configuration to drop, whereas \textsc{SABER} falls far less sharply. By predicting which requests cannot meet their SLA and moving them to the back of the queue, it preserves throughput for the remainder and keeps the system from thrashing.

That advantage widens further under heavy saturation (20 RPS). In workload 1, \textsc{SABER} retains a 8.0\%. lead; in the light-task workload it is 7.0\%.; and in the balanced mix the gap reaches 26\%. In short, \textsc{SABER} behaves like the optimal static setting when capacity is ample, but once contention appears it reallocates GPU time away from hopeless requests and degrades gracefully.

\textbf{Request completion times under \textsc{SABER} exhibit significantly less variation compared to static configurations.}

Figure~\ref{fig:RQ3_stability} plots the mean and standard deviation of ratio of completion time to SLA across different loads. Each curve plots the ratio of actual completion time to SLA threshold shown in the y-axis, while the shaded envelope shows one standard deviation. An adaptive and performant serving configurations should keep the blue/red curves close to or below 1.0 and and keep the shaded band narrow as load rises.

Across all three workloads, the static baseline stays respectable at very light load but becomes erratic as soon as the system nears saturation. 

Across all three workloads, the baseline stays stable at light load but becomes erratic as load increases. In Workload 1, the baseline's  ratio of completion time to SLA rises from 0.58 at 1 RPS to 3 at 20 RPS, far exceeding the SLA threshold of 1.0, while \textsc{SABER} increases more gradually from 0.60 to 1.75. More importantly, the baseline's variance band (shaded region) begins expanding notably from 5 RPS onward, becoming extremely wide by 20 RPS, with its mean exceeding the SLA threshold from 7 RPS onwards;  the coefficient of variation (CV) reaches 52.4\%. \textsc{SABER} maintains a narrower variance band throughout and its CV to 33.7\%, indicating that \textsc{SABER} provides more predictable performance even when the GPU is under pressure.

Workload 2's lightweight tasks do not stress either system, resulting in similar performance with both approaches staying well below the SLA threshold (CV 17.5\% vs 20.8\%). Workload 3 exhibits a distinct pattern with a performance spike at mid-range loads (9-15 RPS). The static baseline's ratio peaks at 1.79 at 15 RPS before declining, while \textsc{SABER} maintains more controlled performance with a maximum ratio of 1.28. The coefficient of variation further confirms \textsc{SABER}'s superior stability: 25.1\% versus the baseline's 36.4\%.
Overall, these results demonstrate that \textsc{SABER}'s adaptive batch size control effectively maintains both performance and predictability across diverse workload patterns, with the advantages becoming more pronounced as system load increases.

\begin{tcolorbox}[colback=white, colframe=black, boxrule=1pt, arc=0pt, left=10pt, right=10pt, top=10pt, bottom=10pt]
\textbf{Findings:} (1) While optimal static configurations are different for each RPS level and thus requires cumbersome reset, \textsc{SABER} consistently maintains high goodput across all loads without reconfiguration. (2) Static configurations exhibit increasingly erratic latency under heavy load, while \textsc{SABER} delivers stable, predictable performance through proactive admission control.

\textbf{Implications:} Production systems often face highly dynamic request loads, varying by an order of magnitude throughout the day. Static configurations cannot accommodate such fluctuations without either degraded performance, resource waste, or constant reset. \textsc{SABER} demonstrates that SLA-aware, adaptive scheduling can sustain high performance across a wide range of load conditions while ensuring predictable latency.
\end{tcolorbox}

%% file: sections/7.discussion.tex
\begin{figure*}[ht]
    \centering
    \begin{subfigure}[t]{0.33\textwidth} 
        \includegraphics[width=\textwidth]{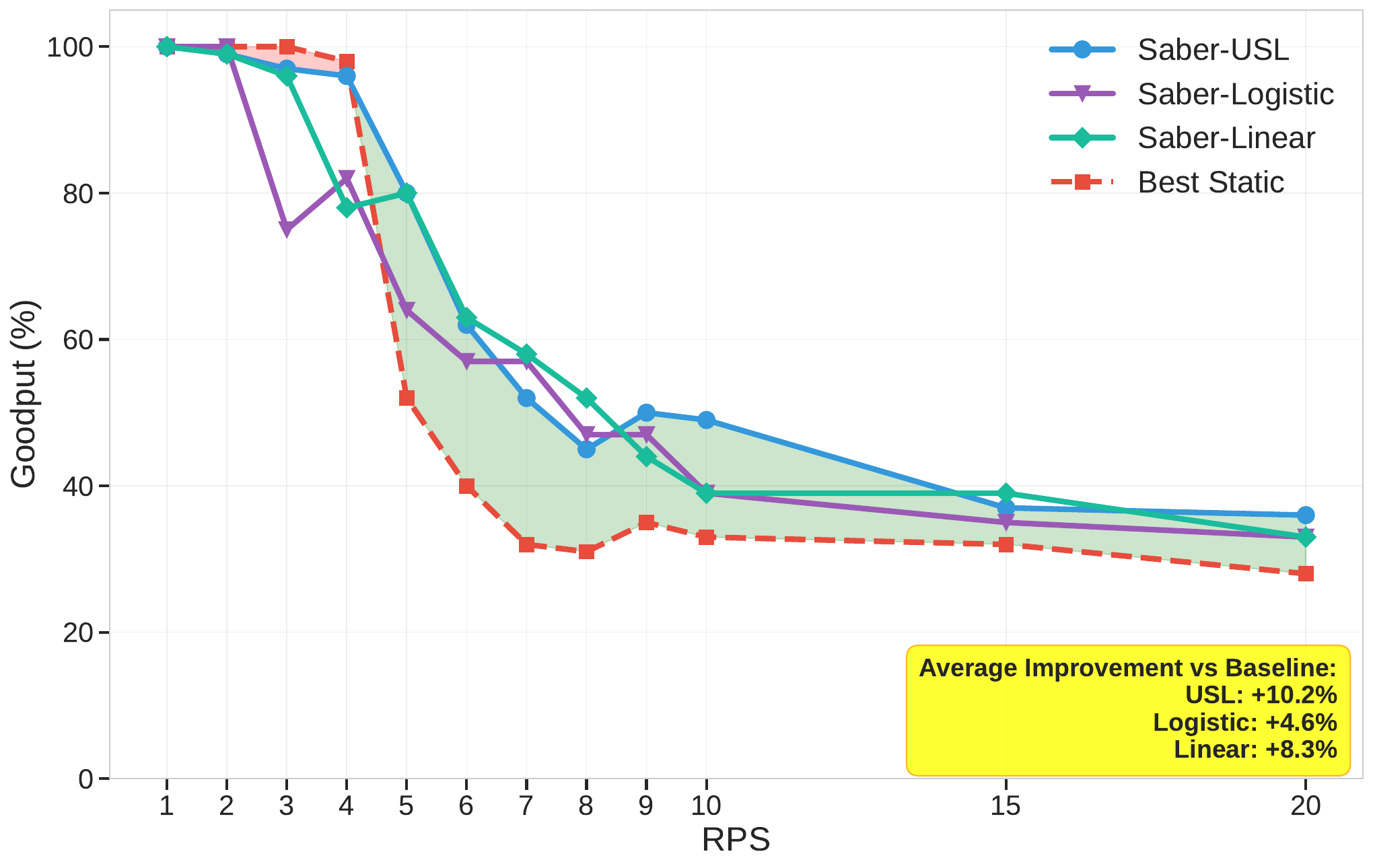}
        \caption{Workload 1}
        \label{fig:est_goodput_workload1}
    \end{subfigure}%
    \hfill
    \begin{subfigure}[t]{0.33\textwidth} 
        \includegraphics[width=\textwidth]{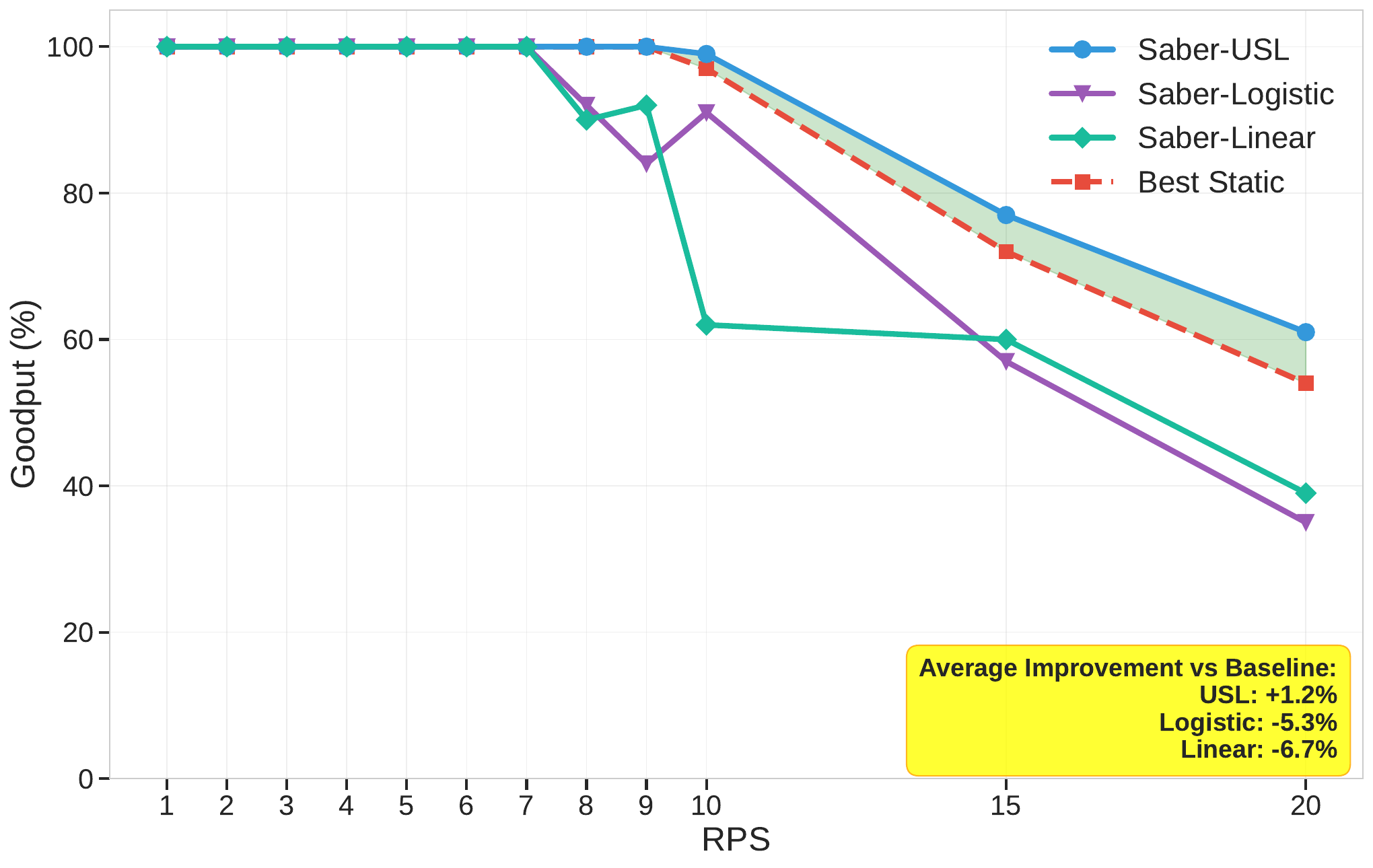}
        \caption{Workload 2}
        \label{fig:est_goodput_workload2}
    \end{subfigure}%
    \hfill
    \begin{subfigure}[t]{0.33\textwidth} 
        \includegraphics[width=\textwidth]{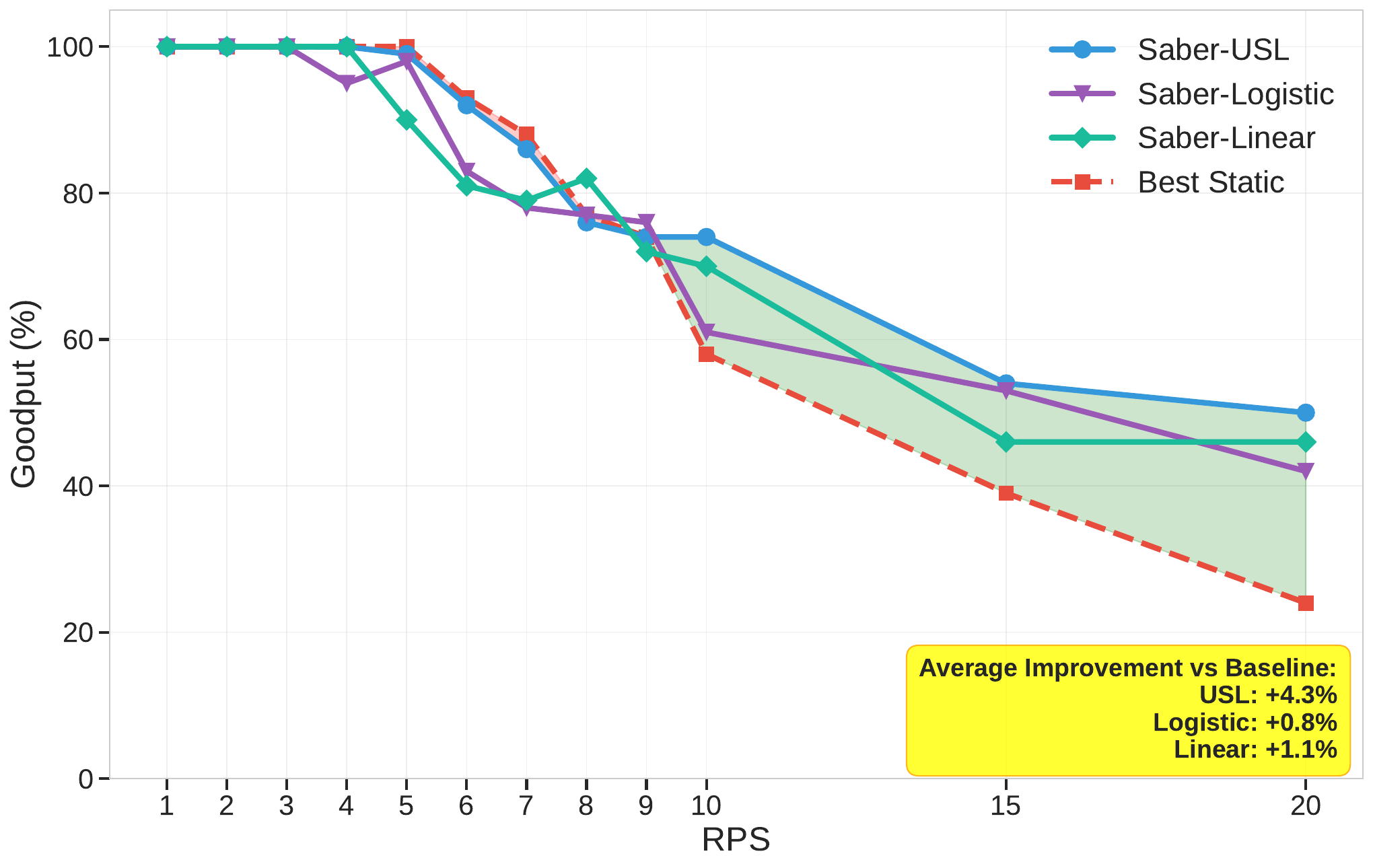}
        \caption{Workload 3}
        \label{fig:est_goodput_workload3}
    \end{subfigure}%
    \caption{Comparative analysis of goodput across four settings: SABER-USL, SABER-Logistic, SABER-Linear, and the best static configuration.}
    \label{fig:est_function_compare}
\end{figure*}

\section{Discussion}
\label{sec:discussion}

Accurate latency prediction lays the foundation of \textsc{SABER}: every admission decision is made by comparing an estimate of a request's completion time against its SLA budget. In this section, we discuss the impact on performance of the core component of~\textsc{SABER}: the estimation function. This is to show that the accuracy of the estimation function affects the performance of~\textsc{SABER}.

To study how estimation quality affects end-to-end performance, we replace the calibrated USL((R\textsuperscript{2}=0.99) estimation function with another two predictors with lower R\textsuperscript{2}: a linear(R\textsuperscript{2}=0.91) and a logistic regression(R\textsuperscript{2}=0.97), while keeping all other logic unchanged. Figure~\ref{fig:est_function_compare} contrasts their goodput against the best static configuration across workloads and request rates.

Across W1 and W3, all \textsc{SABER} variants still outperform the best static configurations. Linear and logistic occasionally outperforming USL around 3-4 RPS due to their conservative admission policies that protect admitted requests. When RPS is low and the response generation is long, the existing requests can finish earlier with less interference and new requests can therefore be processed faster.

The downside of pessimistic prediction becomes obvious in W2(Figures~\ref{fig:est_goodput_workload2}), dominated by short requests whose interference footprint is small. Here capacity is plentiful until very high RPS, so over-conservative admission merely discards viable work.
While USL still yields a modest 1.2\% gain, both linear (-6.7\%) and logistic (-5.3\%)) now fall below the static baseline.

These results highlight high-fidelity prediction, as provided by USL or any equally precise model, is critical for extracting maximum goodput, especially in light-task or mixed workloads where spare capacity exists. Lightweight regressors may suffice when workloads are uniformly heavy, but they risk saturating compute resources as soon as traffic load increases.

%% file: sections/8.theats.tex
\section{Threats to validity}
\label{sec:threats}
In this section, we present the threats to validity.



\subsection{Internal Validity}
Our results rely on accurate estimation of request completion time to make admission decisions. While we use a fitted USL model and evaluate against simpler alternatives, any inaccurate estimation may influence the final outcome. Additionally, our experiments simulate realistic but synthetic workloads; while task compositions come from typical coding scenarios, they might not reflect actual user traces.

\subsection{External Validity}

Our study focuses specifically on decoder-only transformer models (e.g., CodeLLMs) and continuous batching in serving engines like vLLM. While these represent an increasingly common deployment setup, the generalizability of our approach to other model architectures or batching schemes remains unexplored. Likewise, our findings assume a single-node, single-GPU deployment setting. Though this reflects many practical use cases (e.g., indie developers or self-hosted endpoints), large-scale distributed serving systems may face different constraints and require additional coordination mechanisms. The effectiveness of \textsc{SABER} under multi-tenant workloads or with rapidly changing traffic patterns (e.g., bursty spikes) also needs further investigation.

%% file: sections/9.conclusion.tex
\section{Conclusion}
\label{sec:conclusion}

In this study, we have identified a core limitation of continuous batching in LLM serving: static batch size configurations fail to accommodate the diversity of workload compositions and fluctuating request loads, resulting in performance degradation. We propose~\textsc{SABER}, an SLA-aware batching strategy that dynamically adjusts batching decisions using latency estimation. Without requiring reconfiguration or engine modifications,~\textsc{SABER} consistently outperforms the best static settings across varied scenarios—achieving up to 26\% higher goodput and reducing latency variability by up to 45\%. By prioritizing requests most likely to meet their deadlines,~\textsc{SABER} improves system efficiency under contention, offering a practical path toward stable, self-hosted LLM services. As LLMs become integral to AI-assisted software development,~\textsc{SABER} will be essential for unlocking their full potential in resource-constrained environments.

%% file: sections/10.disclaimer.tex
\section{Disclaimer}
\label{sec:disclaimer}
Any opinions, findings, conclusions, or recommendations expressed in this material are those of the author(s) and do noreflect the views of Huawei. Also, ChatGPT-4.0 was used for copy-editing. All experiments, analysis, writing, and results were performed by the authors, who also thoroughly reviewed the final content. This complies with IEEE and ACM policies on AI use in publications.